\documentstyle[preprint,aps]{revtex}
\def\be{\begin{equation}}
\def\ee{\end{equation}}
\begin{document}
\draft
\begin{center}{\bf
TWO-BODY CORRELATIONS AND\\ 
THE UNDERLYING PHASE TRANSITIONS IN BILAYERS}
\end{center}

\begin{center}
{\it 
V.I. Valtchinov $^{(a,b)}$, G. Kalman $^{(a)}$ and K.B. Blagoev $^{(a)}$

$^{(a)}$ Department of Physics, Boston College, Chestnut Hill, MA 02167
, USA\\
$^{(b)}$ Department of Radiology, Brigham and Women's Hospital, \\
Harvard Medical School, Boston, MA 02115, USA\\
\date{\today}
}
\end{center}
\begin{abstract}

The pair correlation functions for a bilayer system of
classical electrons has been calculated through a two-component HNC
formalism. The results show changes in the structure of the correlation
functions as the interlayer distance is increased from zero to about twice
the Wigner-Seitz radius. These changes are indicative of structural phase
transitions which are expected to take place in the crystallized phase and
whose short range orders the liquid phases emulate.
\end{abstract}
\widetext
\section{Introduction}
Electronic bilayer systems ({\it 2D} electron liquids in a neutralizing
background, separated by a distance $d$, each having density $n=1/ \pi a^2$ ,
$a$ being the Wigner-Seitz radius) have attracted a great deal of attention,
both in condensed matter physics \cite{Yang,Narasimhan,Swierkowski} and in
relation to the Penning trapped ion experiments \cite{Gilbert,Dubin}. It has
been realized \cite{Swierkowski,KalmanG} that correlations play a more
important role in such systems than in the {\it 2D} layer or in 
{\it 3D} bulk systems.

Although attempts have been made to estimate the interlayer correlation
functions \cite{Swierkowski,KalmanG,Zheng} no reliable calculation on them
has so far been available. In this paper we report on the results of HNC
calculations of the correlation functions of a classical bilayer for a wide
range of $\Gamma = e^2/akT$ and $d/a$ values. While in most experimental
situations strong magnetic fields are present, and in the semiconductor
bilayers the electronic liquid is degenerate, the classical model is
expected to be quite reliable in the high $\Gamma $ or high $r_{s}$
domain where the electrons are quasi-localized.

\section{Calculation}

The calculation is based on mapping the bilayer into a {\it 2D} two-component
single layer with an interaction matrix \cite{KalmanG}

\begin{eqnarray}
\phi_{11}(r) & = & \phi_{22}(r) = \frac{e^2}{r}\\
\phi_{12}(r) & = & \frac{e^2}{\sqrt{r^2+d^2}}.
\label{eq1} 
\end{eqnarray}
The mapping makes it possible to apply the two-component {\it HNC} formalism,
which is well known \cite{Zhou}. The resulting system of equations is

\begin{eqnarray}
g_{ij}(r) & = & h_{ij}(r) + 1\\
h_{ij}(r) & = & \exp (H_{ij} - \phi_{ij}(r)) - 1\\
H_{ij}(r) & = & h_{ij}(r) - c_{ij}(r)\\
h_{12}(k) & = & \frac{c_{12}(k)}{B(k)}\\
h_{11}(k) & = & \frac{c_{11}(k) + A_{12}(k) h_{12}(k)}{1 - A_{11}(k)}\\
A_{ij}(k) & = & n c_{ij}(k)\\
B(k) & = & [1 - A_{11}(k)]^2 - [A_{12}(k)]^2.
\label{eq2} 
\end{eqnarray}
The system of equations is solved by using Lado's method \cite{Lado} with
the following parameters: maximal radial distance $R\simeq 29$, $\alpha \simeq
2.8$, $N = 500$. Using as convergence monitor the maximum difference between the
input array elements and the corresponding elements after one HNC iteration,
it was found that the difference for the intralayer correlation function
always exceeds the difference for the interlayer function. Therefore, only
the former was monitored, and the iterative solution was obtained for a
maximum difference less than $\varepsilon = 5 \times 10^{-5}$.

\section{Results}

Our results for $g_{11}(r)$ and $g_{12}(r)$ are summarized in Figs. 1,2 and
3,4,5. For low $\Gamma $ values the most interesting effect is the
generation of a partially filled correlation hole in layer $2$ above a
particle in layer $1$. The correlation hole empties both with increasing 
$\Gamma $ and decreasing $d$ (see Figs. 1,2 and 7).

Qualitatively new effects appear for high $\Gamma $ values, where the
quasilocalization of the particles becomes important. It has been proposed 
\cite{Narasimhan,Dubin} that in the crystalline phase (for a monolayer 
$\Gamma_{crit}=137$; for a bilayer the $\Gamma_{crit}(d)$ has not been
established, but probably is of the same order of magnitude; see comment
below) a number of structural phase transitions take place as $d/a$ changed.
Here in the {\it liquid phase} one sees a rather dramatic manifestation of
the change of short-layer order which reflects the underlying structural
phase transitions, as $d/a$ is varied from $0$ to $2$.

At $d=0$ (when the two layers collapse into one single 2D layer) the two
correlation functions $g_{11}(r)$ and $g_{12}(r)$ are identical, since the
two ''species'' are now one and the same and the correlation function is
that of a 2D liquid with double density ($2n$). In the underlying
crystalline phase ''species 1'' and ''species 2'' particles are randomly
distributed on the vertices of a {\it triangular lattice} with lattice
constant $a_{\triangle }=1.35a$. As $d/a$ is increased $g_{11}(r)$ and
$g_{12}(r)$ remain similar, but develop different amplitudes,
$g_{11}(r) < g_{12}(r)$. This reflects the change of the first and second
coordination numbers $q_{11}^{(1)}$, $q_{12}^{(1)}$ and $q_{11}^{(2)}$, 
$q_{12}^{(2)}$ (estimated here from the peak values of $rg_{11}(r)$ and 
$rg_{12}(r)$) and a tendency of the two species to occupy alternating rows of
the triangular lattice which leads to a {\it centered rectangular lattice}
(Fig. 9). This transition seems to be completed around $d/a=0.2$ where the
position of the first peak of $g_{11}(r)$ and that of $g_{12}(r)$ begin to 
separate (Fig. 3), marking the gradual transformation of the rectangular
structure into a {\it centered square lattice} (Fig. 9). At $d/a=0.65$ the
first and the second peaks of $g_{11}(r)$ ($R_{11}^{(1)}$ and 
$R_{11}^{(2)}$) merge (Figs. 4 and 6) which we can interpret as 
the formation of the
latter structure. During these processes the $q_{12}^{(1)}/q_{11}^{(1)}$
ratio changes from $1$ (at $d=0$) to its maximum $1.66$ (at $d/a=0.4$) (Fig.
7), qualitatively corroborating the above picture.The abrupt formation of a
distant second shell at $d/a=0.65$ is indicative of an underlying
(equilateral) square or rhombic structure. Indeed, for $d/a > 0.65$ the
transformation into a {\it centered rhombic lattice} (Fig. 9) can be seen by
the shrinking of $R_{12}^{(1)}$ (Fig. 6). By $d/a=1.5$ the rhombic structure
transforms into a {\it centered triangular lattice} (Fig. 9) where the
particles of the second layer sit over the holes created in the first layer.
This is indicated by the reversal of the decrease of $R_{12}^{(1)}$ (as
particle $2$ shifts from the center of the rhombus into the center of the
triangle, and the settling of $R_{11}^{(1)}$, $R_{11}^{(2)}$ and 
$R_{12}^{(1)}$ near their expected values $a_{\triangle }$, 
$\sqrt{3}a_{\triangle }$ and $\frac{\sqrt{3}}3a_{\triangle }$ , 
with $a_{\triangle}=1.90a$, respectively. Although the staggered 
triangular is the final
equilibrium structure for $d/a>1.5$, further increase of $d$ weakens the
correlations between the layers to the point that by $d/a=3$ the interlayer
correlations practically vanish (Fig. 8).

The $d/a$ values of the structural phase transitions predicted from the
comparison of the respective Madelung energies \cite{Narasimhan,Lado} show
reasonable agreement with our results. However, it should be emphasized,
that in the considered $\Gamma $-range system is liquid and the various
lattice sites are not actual positions of the electrons, but rather
positions of local energy minima, and indicative of the positions of the
succesive high probability ''shells'' in the liquid phase.

Finally, one may wonder whether the bilayer system crystalizes at a lower 
$\Gamma_{crit}$ value than a single {\it 2D} layer, as conjectured by Ref. 3.
Although no definitive answer can be given on the basis of the calculations
performed here, an inspection of Figs. 1 and 2 shows that there is certainly
no tendency for the short-range order to set on a lower $\Gamma $-values
than in the case of a {\it 2D} layer.

(This work has been partially supported by NSF Grant PHY--9115714.)

\figure{
 Fig.1 Intralayer correlation function for different $\Gamma$ values at $d/a = 1$.
 }
\figure{
Fig. 2 Interlayer correlation function for different $\Gamma$ values at $d/a = 1$.
}
\figure{
Fig. 3 Intralayer and interlayer correlation functions. 
}

\figure{
Fig. 4 Intralayer and interlayer correlation functions. 
}

\figure{
Fig. 5 Intralayer and interlayer correlation functions. 
}

\figure{
Fig. 6 Positions of the first and second peaks of the intralayer and interlayer 
correlation functions as indicators of the nearest neighbour positions 
}

\figure{
Fig. 7 Normalized values of the first and second peak amplitudes of the 
interlayer and intralayer correlation functions as indicators of the 
first and second coordination numbers, $q_{11}^{(i)}$ and 
$q_{12}^{(i)}$. 
}

\figure{
Fig. 8 The value of the interlayer correlation function at $r = 0$ for different 
$\Gamma$ values, indicating the degree of correlation between the layers.
}

\figure{
Fig. 9 The four principal lattice structures.
}
\end{document}